\documentstyle[12pt,aasms4]{article}

\def\kms{km\thinspace s$^{-1}$}
\def\cm2{cm$^{-2}$}

\def\eg{{\it e.g.,\/\ }}
\def\sun{_\odot}

\begin{document}

\title{$^{12}$CO J=1-0 Observations of Individual Giant Molecular
Clouds in M81}
\author{Christopher L. Taylor}
\affil{McMaster University and Ruhr-Universit\"at Bochum \\[0.2em]
Astronimisches Institut, Ruhr-Universit\"at Bochum, 
\\[0.2em] Universit\"atsstr 150, D-44780 Bochum, Germany}
\authoraddr{Astronimisches Institut, Ruhr-Universit\"at Bochum, 
Universit\"atsstr 150, D-44780 Bochum, Germany}

\author{Christine D. Wilson}
\affil{McMaster University \\[0.2em]
Department of Physics and Astronomy, Hamilton, Ontario,
Canada, L8S 4M1}
\authoraddr{Department of Physics and Astronomy, Hamilton, Ontario,
Canada, L8S 4M1}

\begin{abstract}

We present $^{12}$CO J=1-0 observations from the Caltech Millimeter Array 
of a field in the nearby spiral galaxy M81.  We detect emission from 
three features that are the size of large giant molecular clouds (GMCs) 
in the Milky Way Galaxy and M31, but are larger than any known in M33 or 
the SMC.  The M81 clouds have diameters $\sim$ 100 pc and molecular masses
$\sim$ 3~$\times$~10$^5$ M$\sun$.  These are the first GMCs to be detected
in such an early type galaxy (Sab) or in a normal galaxy outside the Local 
Group. The clouds we have detected do not obey the size--linewidth relation 
obeyed by GMCs in our Galaxy and in M33, and some of them may be GMC 
complexes that contain several small GMCs.  One of these does show signs 
of sub-structure, and is shaped like a ring section with three separate 
peaks.  At the center of this ring section lies a giant HII region, which 
may be associated with the molecular clouds.

\end{abstract}

\keywords{HII regions -- galaxies: individual (M81) -- galaxies: ISM -- 
ISM: molecules}
 
\section{Introduction}

Giant Molecular Clouds (GMCs) play an important role in the evolution of 
a galaxy.  They are the sites of star formation, and thereby influence 
the evolution of a galaxy's stellar populations.  The stars created in 
them can in turn influence the interstellar medium (ISM) through stellar 
winds and supernovae.  GMCs are the dominant form of the molecular ISM, 
and often contain most of the molecular material in a galaxy.  The global 
properties of GMCs in the Milky Way Galaxy are well known (Sanders, 
Scoville \& Solomon \markcite{SSS}1985; Digel, Bally \& Thaddeus 
\markcite{DBT}1990; Sodroski \markcite{S91}1991).  They are the most 
massive objects in the Galaxy, with masses in the range of 10$^4$ 
to 10$^6$ M$\sun$ and diameters from a few tens of parsecs to $\sim$ 
100 pc.  Hundreds of GMCs have been identified and studied in the Galaxy 
and a few of these are now beginning to be studied in detail via high 
resolution observations (\eg Carpenter, Snell \& Schloerb 
\markcite{CSS}1995).

The properties of GMCs in external galaxies are less well known.  The
Magellanic Clouds have been surveyed and individual GMCs resolved using 
the SEST telescope (e.g. Kutner et al. \markcite{K97}1997; Rubio et al. 
\markcite{RLB}1993).  For other Local Group galaxies, interferometers have 
been used (M31: Vogel et al. \markcite{VBB}1987, Wilson \& Rudolph 
\markcite{WRu}1993; M33: Wilson \& Scoville \markcite{WS}1990; IC 10: 
Wilson \& Reid \markcite{WRe}1991, Wilson \markcite{W95}1995; NGC 6822: 
Wilson \markcite{W94}1994; and NGC 205: Young \& Lo \markcite{YL}1996).  
Interferometric observations of the nearby starburst galaxy M82 have
also identified GMC-like objects (Brouillet \& Schilke \markcite{bs93}
1993; Shen \& Lo \markcite{sl95}1995).  With the exception of the clouds 
in M82, which are bright and massive and more similar to clouds in the 
Galactic Center, the GMCs in these galaxies are quite similar to those 
in the Milky Way Galaxy. However, the clouds in the SMC tend to be smaller 
than those in the Galaxy.  In addition, M33 appears to lack clouds with 
masses $>$ 4~$\times$~10$^5$ M$\sun$, while in our Galaxy the clouds have 
masses up to a few $\times$ 10$^6$ M$\sun$.  Except for these small 
differences, the GMCs in the various galaxies of the Local Group are very
similar, despite enormous variations in the properties of the galaxies 
themselves, which range from early type spirals (M31, an Sb) to late type 
spirals (M33, an Scd), and from dwarf irregulars (IC 10, NGC  6822) to 
dwarf ellipticals (NGC 205). These observations 
suggest that the requirement of 
self-gravitation is more important in determining the properties
of GMCs than is their environment.

To add to the list of galaxies with observations of individual
GMCs, we have obtained CO observations of the nearby Sab galaxy M81.
These observations are the first of such an early type galaxy, as
well as the first to resolve GMCs in a normal
galaxy outside the Local Group.  M81 is also
interesting for this purpose because it has a strong radial metallicity 
gradient (Zaritsky et al. \markcite{ZKH}1994) and thus, observations of 
several fields will allow a comparison of GMCs at different metal 
abundances to determine the effects of varying metal abundance on their 
global properties.  This letter reports on our first detections
of GMCs in one field in M81; the results of a larger, more
sensitive survey currently in progress will be presented in a future paper.

\section{Observations and Data Reduction}

The Owens Valley Millimeter Wave Interferometer was used to observe a
field in M81 in the $^{12}$CO J=1-0 line between 1997 February 19 and May 
30. The field center was  $\alpha$(1950) = 09$^h$ 51$^m$ 32.$^s$9, 
$\delta$(1950) = 69$\arcdeg$ 14$\arcmin$ 21$\arcsec$, a position 
previously detected in the same line by Brouillet et al. 
\markcite{Bet91}(1991) using the NRAO 12-m telescope.  The correlator was 
configured to provide two bandpasses simultaneously, each with 64 channels 
of 0.5 MHz (1.3 \kms) width.  One bandpass was centered on the peak 
velocity of the single dish detection, while the other was set adjacent 
to it to give a combined bandpass of $\sim$ 140 \kms.  Approximately 
8 hours were spent on-source in each of the high and low resolution
configurations.  Flux and gain calibration were obtained by 
observing the nearby quasar 0923+392. This quasar was determined to have 
an average flux of 5.6 Jy from 12 observations in the 3 mm band over the 
period 1996 October 1 to 1997 June 17 for which observations of Neptune
and Uranus were also available. The planets were not available during
our observations themselves; calibration of the flux of 0923+392 relative 
to 3C273 from our data agreed with the fluxes from the OVRO calibrator 
database to within 10\%. To be conservative, we adopt an absolute
uncertainty of 20\% in our absolute flux calibration.

All mapping was performed using the MIRIAD package.  The data were averaged
in velocity to obtain 1 MHz channel maps in the data cube.  The noise in 
the 1 MHz channels of the dirty map was 0.052 Jy beam$^{-1}$.  Cleaned 
maps were made of the bandpass containing the velocity of the single dish 
detection.  The synthesized beam is 3.$\arcsec$2 $\times$ 2.$\arcsec$7, 
with a position angle of $-$24$\arcdeg$.  This beam size corresponds to 
56.3~$\times$~47.5  pc at a distance of 3.63 Mpc (Freedman et al. 
\markcite{f94}1994).  The cleaned data cubes were searched for emission 
from GMCs using the procedures of Wilson \& Scoville \markcite{WS}(1990).  
As a double check, each author conducted a search independently.  Three 
clouds were identified that were common to both lists; in addition, there 
were four very weak features identified by one or the other of us that 
formally met the Wilson \& Scoville \markcite{WS}(1990) criteria, but 
will not be discussed further until follow-up observations are obtained.  
As a further check on the reality of clouds, the bandpass offset from the 
velocity of the single dish detection was mapped and searched for clouds.  
Two very weak features were found, similar to the ones found in the first
bandpass, but nothing as strong as the three main cloud detections.

\section{Results}

The observed and derived properties of the three clouds detected in our
observations are given in Table~\ref{tbl-1}.
 Given in the table are the position
of each cloud, the peak velocity of the line in the frame of the local 
standard of rest, the full width at half maximum velocity width, the
peak brightness temperature of each cloud calculated from the maps with
a conversion of 10.44 K Jy$^{-1}$, the integrated flux density, the
deconvolved cloud diameters along the directions of right ascension and 
declination, the estimated virial mass, and the estimated molecular gas 
mass.  The molecular mass was calculated using $M_{mol} = 1.61\times 10^4 
d_{Mpc}^2 S_{CO}$ M$_\odot$, where $d_{Mpc}$ is the distance in Mpc, 
$S_{CO}$ is the flux density of the cloud in Jy \kms, and a factor of 1.36 
is included to account for the mass of helium in the cloud.  A Galactic 
value of the CO--to--H$_2$ conversion factor was also assumed 
($\alpha=(3\pm 1)\times 10^{20}$ cm$^{-2}$ / (K km s$^{-1}$), 
Strong et al. \markcite{s88}1988; Scoville \& Sanders \markcite{ss87}1987).  
This value is within 15\% of the value suggested by the metallicity--conversion
factor relationship of Wilson \markcite{W95}(1995) for the oxygen abundance
of this region in M81 (Garnett \& Shields \markcite{GS}1987).
Figure~\ref{fig-1} shows the channel maps containing emission from 
the three clouds, while Figure~\ref{fig-2} displays the area-integrated 
spectra for each of the clouds. Figure~\ref{fig-3} shows the integrated 
CO emission for each cloud overlaid on an optical continuum-subtracted 
[OIII] image, and a similar image of the entire field, to show the relative
positions of the clouds.  Note that the figures have not been corrected
for the fall-off in sensitivity at the edges of the primary beam, but 
the flux measurements in Table~\ref{tbl-1} have been corrected
by the appropriate factor. Below we 
discuss each of the clouds individually.

\placetable{tbl-1}

\placefigure{fig-1}
\placefigure{fig-2}

{\it MC~1} (Figure~\ref{fig-3}a) is unresolved by the synthesized beam. This
cloud lies just to the southwest of a bright point source, which is likely
a foreground star. There is no obvious counterpart in the optical image.

{\it MC~2}  (Figure~\ref{fig-3}b) has the shape of a ring section and 
contains three separate peaks, each with roughly the same peak intensity.  
It is unresolved along the width of the ring.  This cloud may in fact 
be a group of 3 clouds, similar to the Orion complex.  The small 
linewidth ($V_{FWHM}$ = 4.7 \kms) suggests that these three objects are 
related, while the small virial mass compared to the molecular mass 
(Table~\ref{tbl-1})
suggests that the entire complex is gravitationally bound.  In the optical 
image  there is an HII region with [OIII] line emission at the location of
the CO emission.  This HII region was also identified by Kaufman
et al. \markcite{KBKH}(1987, their number 172), who measured a 20 cm 
continuum flux density of 0.57 $\pm$ 0.11 mJy, a 6 cm flux density of 
0.52 $\pm$ 0.07 mJy, and a spectral index of $-$0.08 $\pm$ 0.19.
The flat spectral index (consistent with zero) indicates that the radio
continuum flux is entirely thermal in origin, and that there are no
supernovae remnants associated with the HII region.  This could be an 
indication that the HII region is relatively young.  This HII region was 
also one of those used by Garnett \& Shields \markcite{GS}(1987, their 
number 2) in their study of the distribution of metal abundance in M81.  
>From their spectra they estimated the number of ionizing photons per 
second to be 6.4~$\times$~10$^{50}$, the oxygen abundance to be 12 $+$ 
log(O/H) = 8.79, and the electron temperature, T$_e$, to be 7200 K.

{\it MC~3}  (Figure~\ref{fig-3}c) is irregularly shaped and elongated 
along a position angle of $\sim$ 45$\arcdeg$.  Unlike MC~2, it does not 
appear to have any substructure, and is resolved in all directions.  It 
is located just north of MC~1, but is separated in velocity by 8 \kms.
Because of the proximity of MC~3 to MC~1, it is reasonable to consider
whether they are in fact two components of a single cloud.
However, the boundaries of the two clouds are distinct at the 50\% of
peak intensity contour, while  the peaks of the two clouds are separated
by 90 pc in {\it projected} distance, and the extreme edges by nearly 
twice that. These properties, 
together with the combined emission spectrum that 
shows a dip to zero emission between the peak intensities of the two clouds,
strongly suggest that these are two separate clouds.

\section{Discussion}

The field we have observed was observed by Brouillet et al. 
\markcite{Bet91}(1991) with the NRAO 12-m telescope as their field S6, 
for which they reported an integrated intensity, $I_{CO}$, of 0.4 K \kms.  
Assuming a gain of 34 Jy K$^{-1}$, this intensity corresponds to an 
integrated flux density of 13.6 Jy \kms, while we have detected a total of 
7.9 Jy \kms\ in these three clouds.  Clearly we have not detected all the 
flux observed by Brouillet et al. \markcite{Bet91}(1991).
There are two possible explanations for this: 1) our interferometric
observations are missing flux due to missing short baselines in our
$u-v$ coverage, or 2) the missing flux is weak enough to be below our
detection threshold, {\it i.e. in GMCs not massive enough to detect in
our data}.  These explanations are not mutually exclusive;
for example, we could be missing an extended component of molecular gas 
with low column density. The width of the single dish
line is 14.2 \kms, while the spread in velocities of the different
clouds we have detected is 7.8 \kms, which suggests that we are missing
a component of the molecular ISM with a larger velocity dispersion
than the large clouds we have found. This result is similar to what is 
seen in M33, where the single dish line widths are also larger than
the velocity dispersion of the clouds seen interferometrically (Wilson 
\& Scoville \markcite{WS}1990). The missing component could be either 
lower-mass molecular clouds or a population of translucent or 
high-latitude molecular clouds, such as are seen in the Milky Way (i.e. 
Blitz, Magnani, \& Mundy \markcite{bmm84}1984).

One important question is how the properties of the clouds in 
M81 compare to those in other galaxies.  From Table~\ref{tbl-1}, we see that 
the molecular mass of the clouds is on the order of 3~$\times$~10$^5$
M$\sun$. Such masses would not be unusual for clouds in the Milky Way Galaxy 
or M31 (Wilson \& Rudolph \markcite{WRu}1993), but would be at the 
extreme high end of the distribution in M33 (Wilson \& Scoville 
\markcite{WS}1990) and are larger than the cloud masses in the SMC 
(Rubio et al. \markcite{RLB}1993).  There is a similar trend for the 
physical sizes of the GMCs: the M81 clouds have diameters $\sim$ 100 pc, 
similar to the largest GMCs known in the Milky Way (Sodroski 
\markcite{S91}1991), but much larger than
what is seen in M31 (although to date there are only 4 GMCs known in M31).
In M33, the diameters only reach $\sim$ 60 pc, and the SMC clouds are
smaller still.  It is possible that the maximum mass and diameter of 
GMCs depend upon the morphology of the host galaxy, 
such that late-type galaxies have smaller,
less massive GMCs than early-type galaxies.  To prove this conjecture
will require more extensive searches for GMCs in galaxies like M31 and M81,
as well as extending the searches to more galaxies to achieve 
sufficient numbers of galaxies for statistical analyses.

A third property to compare is the linewidth, which is thought to be a 
measure of the clump-clump velocity dispersion within GMCs.  For the
three M81 clouds, the linewidth lies between 4.7 \kms\ and 7.7 \kms, which
is well within the range spanned by the GMCs in M33, M31, IC 10 and the 
SMC.  However, given the large sizes of these clouds, the linewidths are 
somewhat low.  Figure~\ref{fig-4} plots the size-linewidth relation for 
the two resolved clouds from M81 and 18 GMCs from M31, M33, and IC 10.  
The curve drawn in the figure shows the size-linewidth relation fitted 
by Wilson \& Scoville \markcite{WS}(1990) to the GMCs in M33, $V_{FWHM} = 
1.2 D_{pc}^{0.5}$.  Similar curves have been fit to the population of 
GMCs in the Milky Way Galaxy (\eg Sanders et al. \markcite{SSS}1985;
Sodroski \markcite{S91}1991).  
The existence of this relationship has by some authors (\eg Issa, 
MacLaren \& Wolfendale \markcite{IMW}1990) been attributed to be a 
consequence of criteria used to define the borders of a cloud and the 
crowded condition of the GMCs both spatially and in velocity in the 
inner Galaxy.  This argument is not applicable to GMCs in the outer 
Galaxy, or in other galaxies, where the same sort of relationship has 
been found (Sodroski \markcite{S91}1991; Wilson \& Scoville 
\markcite{WS}1990).  
\placefigure{fig-4}

The clouds in M81 do not obey this size-linewidth relationship, with 
linewidths that are too small for their sizes.  Similarly narrow 
linewidths are found by Brouillet et al. \markcite{Bet97}(1997) for six
molecular complexes 
in a different region of M81.  Although the clouds we have discovered
appear to be gravitationally bound (Table~\ref{tbl-1}), they may not be 
individual GMCs.  This possibility is especially true of MC~2, which has 
the shape of a circular arc, and shows three distinct peaks in the spatial 
distribution of its CO emission.  In this case, perhaps the three peaks 
represent GMCs that would obey the size-linewidth relationship if they 
could be observed with high enough spatial resolution.
An analog to this in our own Galaxy may be the Orion molecular complex.
The diameter of this complex is approximately 130 pc, and the molecular
mass inferred from the CO emission is 2.3~$\times$~10$^5$ M$\sun$
(Maddalena et al. \markcite{MMMT}1986).  The composite linewidth for all 
the clouds in the complex, including the neighboring Monoceros R2 complex, 
is 6 \kms~ (Maddalena et al. \markcite{MMMT}1986), similar to that of 
our MC~2.

\section{Summary}

We have reported on $^{12}$CO J=1-0 interferometric observations
of a field in M81 with the Owens Valley Millimeter Array.  With its high
resolution, we have detected three emission features which are roughly 
the size and mass of the largest giant molecular clouds in the Galaxy.
M81 is now the earliest-type and most distant normal galaxy in which GMCs 
have been detected.  One of these clouds is shaped like a circular arc, 
with a giant HII region at the center.  The clouds we have detected, 
although similar to the largest GMCs in M31 and our own Galaxy, are more 
massive and larger than those seen in later-type galaxies such as M33 and 
the SMC.  At least two of the three detected clouds do not obey the 
size-linewidth relationship as delineated by  GMCs in the Milky Way Galaxy 
and M33.  Thus, these objects
may not be individual GMCs, but may instead be complexes of several clouds,
like the Orion complex in our Galaxy.  We plan to extend this work by
observing with greater sensitivity to detect less massive clouds,
and by observing other fields in M81 to compare the properties
of GMCs at different locations within the galaxy. 

\acknowledgments
 
This research is supported through a grant from the Natural Sciences 
and Engineering Research Council of Canada.  Observations with the
OVRO Millimeter-wave Array are supported by U.S. National Science 
Foundation grant AST 93--14079.

\clearpage

\figcaption[fig1.ps]{Channel maps showing the CO emission from
the three M81 GMCs.
The channels are 1 MHz (2.6 \kms) wide,
and the contours  are ($-4,-3,-2,2,3,4) \times 0.052$
Jy beam$^{-1} = 1\sigma$. Negative contours are indicated by dashed
lines and the central velocity of each channel is given in the lower left
corner.
\label{fig-1}}

\figcaption[fig2.ps]{Spectra of the M81 GMCs, obtained by integrating over 
the area of the CO emission through the channels of the data cube.
The velocity of MC 1 is offset by $\sim$ 7 \kms\ from the other clouds,
and is indicated by the arrow.
\label{fig-2}}

\figcaption[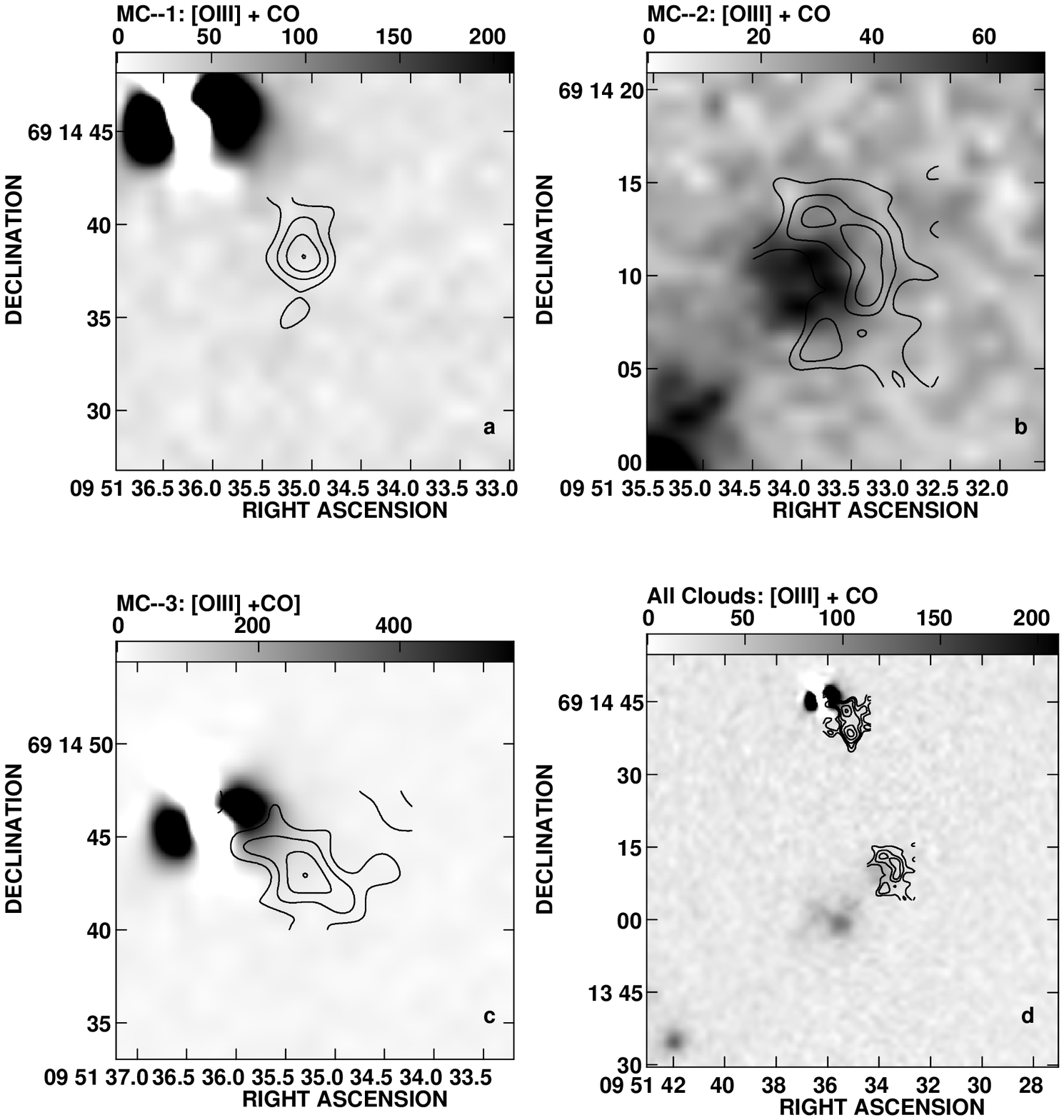]{[OIII] -- CO overlays.  
The optical images in the line and
neighboring continuum were obtained from the NCSA Astronomy Digital 
Image Library (URL: http://imagelib.ncsa.uiuc.edu/imagelib.html).  The
images were registered, the continuum subtracted, and the resulting
image was registered with our CO maps. (a) MC 1. CO contours are
$(1,2,3,4)\times 0.338$ Jy beam$^{-1}$ \kms $ = 1\sigma$ and the map
was made by averaging over 5 MHz.
(b) MC 2. CO contours are 
$(1,2,3)\times 0.2028$ Jy beam$^{-1}$ \kms $ = 1\sigma$ and the
map was made by averaging over 2 MHz.
(c) MC 3. CO contours are 
$(1,2,3,4)\times 0.312$ Jy beam$^{-1}$ \kms $ = 1\sigma$ and the
map was made by averaging over 4 MHz.
(d) The entire field, showing the relative positions of the clouds. 
The CO contours are 
$(1,2,3,5,7)\times 0.2028$ Jy beam$^{-1}$ \kms.
  The southern spiral arm of 
M81 can be traced out along the line of HII regions running from northwest 
to southeast through the image.  
\label{fig-3}}

\figcaption[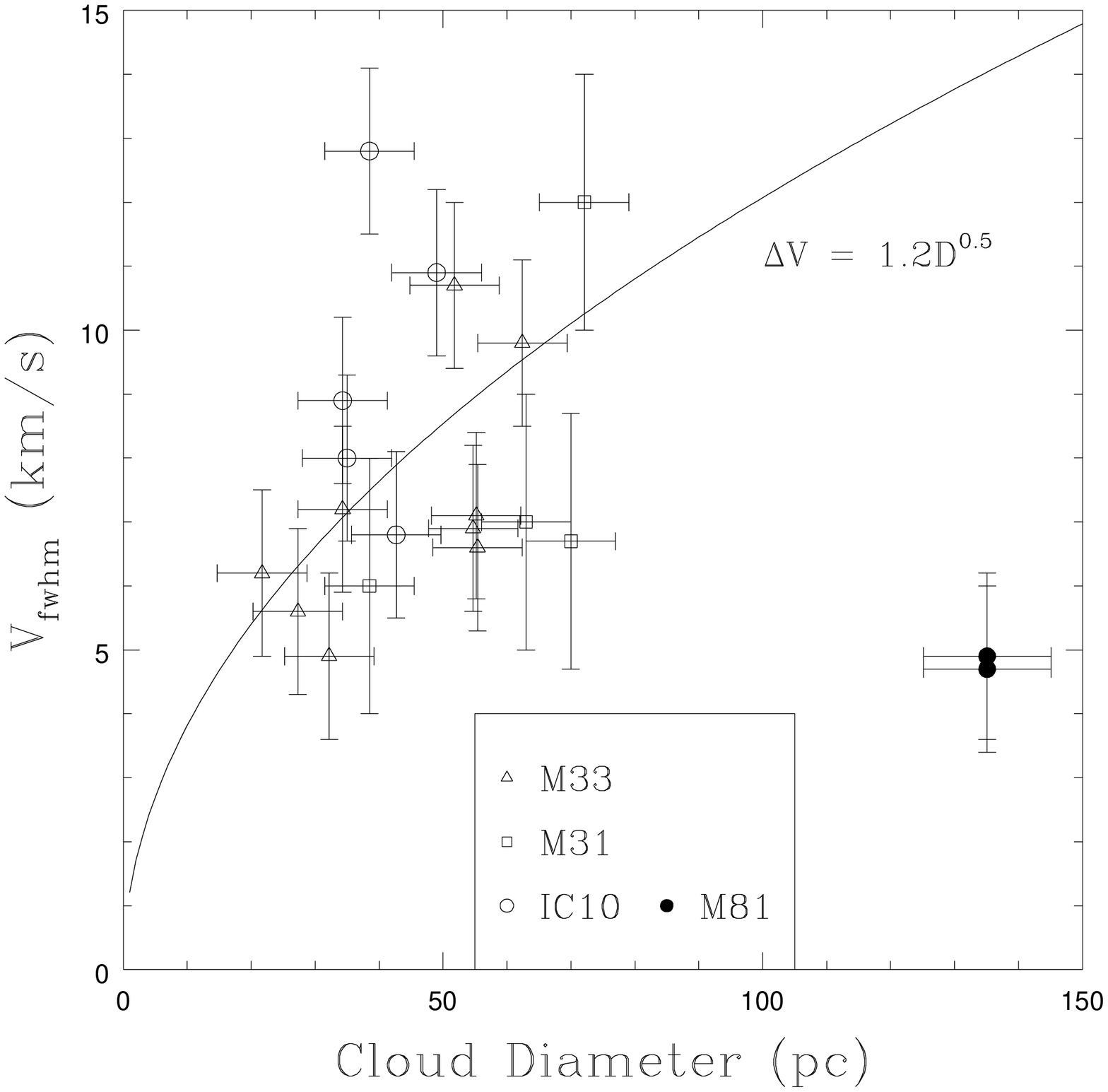]{A plot of line width versus diameter for GMCs in
M33 ({\it triangles}), M31({\it squares}), IC 10 ({\it circles}) and M81
({\it filled circles}).  The curve shows the size--linewidth relation
fit by Wilson \& Scoville (1990) to GMCs in M33.
\label{fig-4}}

\end{document}